\documentclass[%
 reprint,
superscriptaddress,
 amsmath,amssymb,
 aps, prl
]{revtex4-2}

\usepackage{graphicx}
\usepackage{dcolumn}
\usepackage{bm}
\usepackage{xcolor}

\begin{document}

\preprint{APS/123-QED}

\title{Mie-enhanced micro-focused Brillouin light scattering with wavevector resolution}
\author{Jakub Krčma}
\affiliation{%
Faculty of Mechanical Engineering, Institute of Physical Engineering, Brno University of Technology, Technick\'{a} 2, Brno, 616 69, Czech Republic\\
}%
\author{Ond\v{r}ej Wojewoda}
\email{ondrej.wojewoda@vutbr.cz}
\affiliation{%
 CEITEC BUT, Brno University of Technology, Purky\v{n}ova 123, Brno, 612 00, Czech Republic\\
}%
\author{Martin Hrtoň}
\affiliation{Faculty of Mechanical Engineering, Institute of Physical Engineering, Brno University of Technology, Technick\'{a} 2, Brno, 616 69, Czech Republic\\
}
\affiliation{%
 CEITEC BUT, Brno University of Technology, Purky\v{n}ova 123, Brno, 612 00, Czech Republic\\
}%
\author{Jakub Holobr\'{a}dek}
\affiliation{%
 CEITEC BUT, Brno University of Technology, Purky\v{n}ova 123, Brno, 612 00, Czech Republic\\
}%
\author{Jon Ander Arregi}
\affiliation{%
 CEITEC BUT, Brno University of Technology, Purky\v{n}ova 123, Brno, 612 00, Czech Republic\\
}%
\author{Jaganandha Panda}
\affiliation{%
 CEITEC BUT, Brno University of Technology, Purky\v{n}ova 123, Brno, 612 00, Czech Republic\\
}%
\author{Michal Urb\'{a}nek}
\email{michal.urbanek@ceitec.vutbr.cz}
\affiliation{Faculty of Mechanical Engineering, Institute of Physical Engineering, Brno University of Technology, Technick\'{a} 2, Brno, 616 69, Czech Republic\\
}
\affiliation{%
 CEITEC BUT, Brno University of Technology, Purky\v{n}ova 123, Brno, 612 00, Czech Republic\\
}%
\date{\today}

\begin{abstract}
Magnons, the quanta of spin waves, are magnetic excitations of matter spanning through the entire crystal’s Brillouin zone and covering a wide range of frequencies ranging from sub-gigahertz to hundreds of terahertz. Magnons play a crucial role in many condensed matter phenomena, such as the reduction of saturation magnetization with increasing temperature or Bose-Einstein condensation. However, current experimental techniques cannot resolve magnons with wavevectors between 30 and 300\,rad\,{\textmu}m$^{-1}$. In this letter, we address this gap by tailoring the light in Brillouin light scattering process with dielectric periodic nanoresonators and thus gaining access to the previously unmeasurable spin waves with full wavevector resolution using table-top optical setup. Filling this gap can stimulate further experimental investigations of the fundamental phenomena associated with magnons but also stimulate the application of magnonics in computational and microwave devices. In addition, the same methodology can be applied to other excitations of matter, such as phonons, opening up new possibilities in e.g. mechanobiological studies.
\end{abstract}

\maketitle

\begin{figure*}
\includegraphics{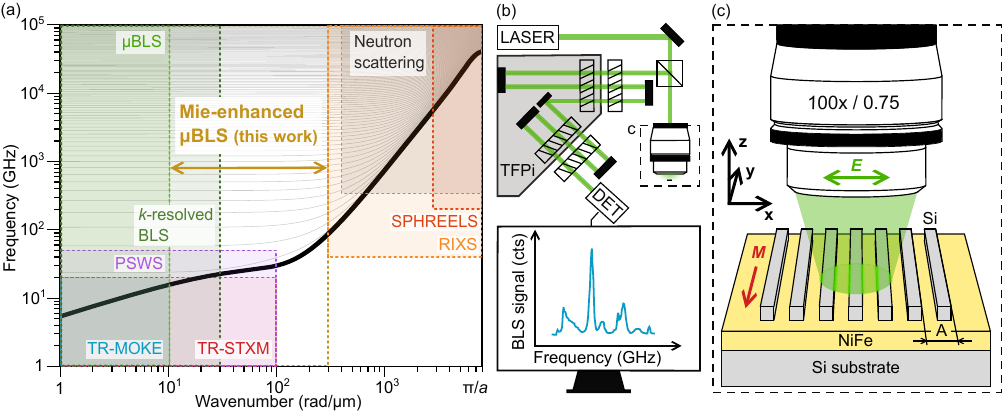}
\caption{\label{fig:Fig1} Comparison between experimental techniques, and schematics of Mie-BLS measurement. (a) Schematics of a spin wave dispersion relation in a permalloy thin film spanning over the whole Brillouin zone in an external magnetic field of 50\,mT. Various experimental techniques are highlighted with their corresponding accessible frequency and wavenumber ranges. The range achieved using Mie-enhanced \textmu BLS in this work is emphasized. (b) Experimental setup for \textmu BLS measurements. Green laser light (532\,nm) is guided and focused onto the sample through an objective lens ($N\!A=0.75$). The backscattered light from the sample is collected by the same lens and analyzed using a tandem Fabry-Perot interferometer to measure the spin wave frequency spectrum. (c) Sample and measurement geometry. The sample consists of silicon nanoresonators arranged with a periodicity $A$ on top of a permalloy 27.6\,nm-thick film deposited onto a silicon substrate.}
\end{figure*}

The field of magnonics has emerged as a promising area of research, focusing on the collective magnetic excitations known as spin waves and their associated quasi-particles, magnons. Magnonics can be categorized into two areas of interest: applied/microscopic magnonics and research of fundamental properties of matter. The field of applied/microscopic magnonics searches for new energy efficient computation paradigms, devices, or study interesting behavior of matter at micro/nanoscale \cite{wang_nanoscale_2024, chumak_magnon_2015, wojewoda_propagation_2020, flebus_recent_2023, klima_zero-field_2024, branford_materials_2024, wojewoda_unidirectional_2024}. These studies are usually performed on spin waves with their wavelength spanning only from the center of Brillouin zone to approx. 100\,rad\,{\textmu}m$^{-1}$, and utilize variety of magneto-optical techniques such as \textit{k}-resolved Brillouin light scattering (\textit{k}-resolved BLS), micro-focused BLS (\textmu BLS), time-resolved magneto-optical Kerr effect (TR-MOKE), time-resolved scanning transmission x-ray microscopy (TR-STXM), or propagating spin-wave spectroscopy (PSWS) \cite{sebastian_micro-focused_2015, madami_application_2012, wojewoda_modeling_2024, flajsman_wideband_2022, ogasawara_time-resolved_2023, stigloher_snells_2016, qin_nanoscale_2021, sluka_emission_2019, vanatka_spin-wave_2021, devolder_propagating-spin-wave_2023}. In contrast, to investigate fundamental properties of matter, such as chiral splitting in altermagnets \cite{smejkal_chiral_2023,liu_chiral_2024} or spin-orbital separation \cite{schlappa_spinorbital_2012} one usually needs to investigate magnons across the whole Brillouin zone and at terahertz frequencies. This research utilizes inelastic neutron scattering, resonant inelastic X-ray scattering (RIXS), or spin-polarized high-resolution electron energy-loss spectroscopy (SPHREELS) \cite{ament_resonant_2011, mook_neutron-scattering_1985, plant_spinwave_1977, zakeri_giant_2024}. However, there is no single experimental technique that covers the whole range of interest, as each technique is limited by technical or fundamental constraints, see Fig.~\ref{fig:Fig1}(a).   

In wavenumber, imaging techniques are limited by the spot size to approx. 10\,rad\,{\textmu}m$^{-1}$ for \textmu BLS and TR-MOKE and 100\,rad\,{\textmu}m$^{-1}$ for TR-STXM. The \textit{k}-resolved BLS is limited by the conservation of momentum to approx. 30\,rad\,{\textmu}m$^{-1}$ \cite{sebastian_micro-focused_2015} and the PSWS by the excitation efficiency to approx. 100\,rad\,{\textmu}m$^{-1}$ \cite{liu_long-distance_2018, yu_omnidirectional_2013, baumgaertl_nanoimaging_2020}. In frequency, \textmu BLS has no limitation in the range of interest as various types of spectrometers can be used, such as (tandem) Fabry-Perot interferometers \cite{lindsay_construction_1981}, virtual phase arrays \cite{chagnon-lessard_high-contrast_2024}, or grating monochromators  \cite{scarponi_high-performance_2017}. Contrary to that, techniques which rely on coherent excitation (pump-probe, electrical detection) are limited by the used electronics, typically to tenths of gigahertz. Techniques used to measure spin waves across the whole Brillouin zone (inelastic neutron scattering, SPHREELS, RIXS) are limited in resolving lower wavenumbers and frequencies by their resolution (typically over 300\,rad\,{\textmu}m$^{-1}$ and 100\,GHz).

In between these two areas, a relevant experimental blind spot becomes apparent, which spans approximately from 100\,rad\,{\textmu}m$^{-1}$ to 300\,rad\,{\textmu}m$^{-1}$ for low frequencies, and from 30\,rad\,{\textmu}m$^{-1}$ to 300\,rad\,{\textmu}m$^{-1}$ for high frequencies, see Fig.~\ref{fig:Fig1}(a). Various attempts to fill this gap were made by utilizing near-field enhancement of the \textmu BLS signal, which can lift the restriction posed by the conservation of momentum law by introducing an imaginary part to the momentum of the incident electro-magnetic wave. This principle was used to enhance the wavenumber range by utilizing plasmon resonances, however in practical applications the signal yield is very low \cite{jersch_mapping_2010, freeman_brillouin_2020, lozovski_plasmon-enhanced_2024}. Recently, it was reported that the dielectric nanoresonators (Mie resonances) enhance both the wavevector sensitivity and the signal yield \cite{wojewoda_observing_2023, wojewoda_phase-resolved_2023}. However, without coherent excitation, individual nanoresonators lack wavevector resolution, and thus can only provide limited information about the studied excitation.

In this letter, we address these limitations by introducing periodic dielectric nanostripes that provide full wavevector resolution, including its magnitude and direction, across the previously inaccessible range. This is achieved by superimposing the wave vector of light and the lattice vector of the periodic nanostripes. This approach draws an inspiration from propagation of waves in periodic potentials, such as spin waves in artificial magnonic crystals \cite{Chumak_2017}.

To measure inelastically scattered spectra, we use a standard \textmu BLS setup, see Fig.~\ref{fig:Fig1}(b), utilizing a single-mode laser (COBOLT Samba) with a wavelength of 532\,nm. The laser light is focused by a high-numerical-aperture objective lens ($N\!A = 0.75$) onto a 440\,nm spot on the sample \cite{wojewoda_observing_2023}. The backscattered light is then collected by the objective lens and analyzed by the multi-pass tandem Fabry-Perot interferometer, which measures the spin wave frequency spectrum \cite{lindsay_construction_1981}. 

The sample consists of a $(27.6\pm{0.1})\,\mathrm{nm}$-thick permalloy film deposited on a silicon substrate (for thickness measurement details see End Matter). Periodic silicon nanostripes, acting as Mie resonators, were fabricated on top of the magnetic film by electron beam lithography, ion beam sputtering, and a lift-off process. The stripes form gratings with varying periodicities $A$, each grating covering $6.5\times{6.5}\,\text{\textmu m}^2$ of the sample, see Fig.~\ref{fig:Fig1}(c).

\begin{figure}
\includegraphics{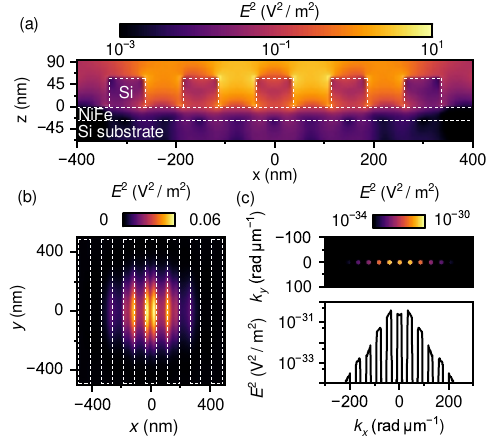}
\caption{\label{fig:Fig2} Finite-difference time-domain (FDTD) simulations of the electric field distribution of the incident laser light on the magnetic layer. (a) Logarithmic-scale electric field distribution in the $x,z$-plane, showing a cross-sectional view of the dielectric grating. Strong electric field localization is observed at the edges of the stripes (hot spots). White dotted lines indicate the interfaces. (b) Electric field distribution in the $x,y$-plane at the middle of the permalloy layer. Periodic field modulation is visible within the Gaussian beam spot. White dotted lines indicate the stripe positions. (c) Reciprocal electric field distribution in the $x,y$-plane, obtained via Fourier transformation of the data from panel (b). The 1D plot (lower panel) shows the logarithmic intensity profile at $k_y=0$, highlighting the field's periodic distribution in the reciprocal space.}
\end{figure}

To quantitatively describe the Mie-BLS process, it is essential to understand how the incident light interacts with the dielectric nanoresonators and the magnetic material. When the driving incident electric field (${E}_{\mathrm{dr}}$) interacts with the periodic dielectric nanoresonator array with the lattice constant $A$, it generates a strongly localized electric field (hot spots) at the edges of the individual nanoresonator stripes within the illumination Gaussian spot, see Fig.~\ref{fig:Fig2}(a). These hot spots are the result of excitation of Mie resonances within the nanoresonators, which modulates the field distribution inside the magnetic layer [Fig.~\ref{fig:Fig2}(b)]. This modulation leads to a periodic electric field distribution, enabling the light to couple only to states with wavenumbers corresponding to multiples of $2\pi/A$, even beyond the free-light accessible boundaries given by momentum conservation [Fig.~\ref{fig:Fig2}(c)] \cite{freeman_brillouin_2020, wojewoda_observing_2023}. At the same time, the high symmetry of periodic dielectric nanoresonators in the \textit{y}-direction [see Fig.~\ref{fig:Fig2}(c)] leads to a strong localization of the electric field along the $k_{y}$-direction in the reciprocal space. The widths of the $k=0$ peak and all other higher order peaks are determined by the width of the illumination laser spot.

This modulated electric field interacts inelastically with the dynamic magnetization (spin waves) via magneto-optical coupling, inducing polarization
\begin{equation} \label{eq:Polarization}
\boldsymbol{P}_{\mathrm{m}}(\boldsymbol{r},\omega+\omega_{\mathrm{m}})=\boldsymbol{\chi}_{\mathrm{m} }(\boldsymbol{r},\omega_{\mathrm{m}}) \boldsymbol{E}_{\mathrm{dr}}(\boldsymbol{r},\omega),
\end{equation}
where $\boldsymbol{\chi}_{\mathrm{m} }(\boldsymbol{r},\omega_{\mathrm{m}})$ is the dynamic magnetic susceptibility tensor \cite{qiu_surface_2000}, $\boldsymbol{E}_{\mathrm{dr}}(\boldsymbol{r},\omega)$ the field generated in the layer by the incident laser beam (driving field), and $\omega_{\mathrm{m}}$ the frequency of spin waves. Since the frequency of the incoming radiation is much larger than that of the spin waves, we omit its dependency on the induced polarization.

The polarization acts as a local source of radiation and its out-coupling from the magnetic layer and subsequent propagation towards the detector can be described using the dyadic Green’s function formalism \cite{wojewoda_modeling_2024}. However, due to the broken spatial symmetry caused by the presence of the dielectric nanoresonator array, the dyadic Green’s function is not known analytically and it needs to be derived from numerical simulations. Recalling that it represents the impulse response of the system to a point source, the task of finding the dyadic Green’s function amounts to running a large set of simulations, where the position of the point source is varied within a sufficient range and with an adequate resolution. Alternatively, one can switch to reciprocal space and attempt to reconstruct the angular spectrum representation of the dyadic Green’s function \cite{wojewoda_observing_2023}, i.e. an impulse response to a plane wave, but since the dielectric nanostripes do not possess radial symmetry, the number of simulations to obtain the dyadic Green's function with a sufficient precision becomes prohibitively high.

To overcome these limitations, we exploited the reciprocity theorem of electromagnetism, drawing an inspiration from a similar approach for the scattering process in scanning near-field optical microscopy (s-SNOM) \cite{neuman_mapping_2015}. The reciprocity theorem simplifies the scattering problem by introducing the concept of a virtual source located at the detector and providing a clear mathematical link between the electric field generated by this virtual source and the radiation emitted by an actual source, later collected by the detector. In accordance with our BLS detection scheme, the virtual source is represented by an electric dipole $\boldsymbol{p}_{\mathrm{v}}$ oriented perpendicularly to the incident laser light. Denoting $\boldsymbol{E}_{\mathrm{v}}(\boldsymbol{r})$ the electric field generated by the virtual source within the magnetic layer and $\boldsymbol{E}_{\mathrm{m}}(\boldsymbol{r}_{\mathrm{det} })$ the electric field produced by the polarization source $\boldsymbol{P}_{\mathrm{m}}(\boldsymbol{r})$ at the position of the detector $\boldsymbol{r}_{\mathrm{det} }$, the reciprocity theorem enforces the following relation
\begin{equation} 
\boldsymbol{p}_{\mathrm{v} } (\boldsymbol{r}_{\mathrm{det} }) \cdot \boldsymbol{E}_{\mathrm{m} } (\boldsymbol{r}_{\mathrm{det} })  = \int \mathrm{d}r^{3} \, \boldsymbol{P}_{\mathrm{m} } (\boldsymbol{r}) \cdot \boldsymbol{E}_{\mathrm{v} } (\boldsymbol{r}),
\end{equation}
where the integration spans, in principle, the entire magnetic layer, but in practice is limited to the illumination spot. Assuming that the induced polarization has the form given by Eq.~\ref{eq:Polarization} and that only a single magnon with a lateral wavevector $\boldsymbol{k}_{\mathrm{m} }\!\perp\! z$ is being excited (for details see End Matter), the detected BLS signal can be expressed as

\begin{align}
\label{eq:ReciTheorem}
\sigma( & \boldsymbol{k}_{\mathrm{m} }, \omega_{\mathrm{m}} )  \sim \vert \boldsymbol{p}_{\mathrm{v} } (\boldsymbol{r}_{\mathrm{det} }) \cdot \boldsymbol{E}_{\mathrm{m} } (\boldsymbol{r}_{\mathrm{det} }) \vert^{2} = \notag \\ & = \Big\vert \sum\limits_{i,j} \int \mathrm{d}z \, {\chi}_{\mathrm{m} }^{ij} (z,\omega_{\mathrm{m} }) \, T_{ij}(\boldsymbol{k}_{\mathrm{m} } , z) \Big\vert^{2},
\end{align}
where the transfer function $T_{ij}(\boldsymbol{k}_{\mathrm{m} } , z)$ defined as
\begin{equation}
T_{ij}(\boldsymbol{k}_{\mathrm{m} } , z) = \int \mathrm{d}r_{\parallel}^{2} \, {E}_{\mathrm{v} }^{i} (\boldsymbol{r}) {E}_{\mathrm{dr} }^{j} (\boldsymbol{r}) \, e^{i \boldsymbol{k}_{\mathrm{m} } \cdot \boldsymbol{r}_{\parallel} } 
\end{equation}
effectively determines the range of spin wave wavevectors that the system is able to couple with, regardless of whether these spin wave states are occupied or not. Note that the vector nature of the electric fields makes the transfer function a rank two tensor. Furthermore, the result of the integration over the vertical coordinate $z$ will generally depend on the type of magnon that is being excited, but given the relatively large thickness of our metallic layer with respect to the penetration depth of the electric field---so that it effectively probes only the top portion of the magnetic layer---we simplified our analysis by recording the field and evaluating the transfer function only within a single plane. 

As Eq.~\ref{eq:ReciTheorem} suggests, the problem is reduced to finding the driving field ($\boldsymbol{E}_{\mathrm{dr}}$), the virtual field ($\boldsymbol{E}_{\mathrm{v}}$), and the dynamic magnetic susceptibility ($\boldsymbol{\chi}_{\mathrm{m} }$). The electric fields are obtained by performing two finite-difference time-domain (FDTD) simulations with polarizations perpendicular to each other, and the dynamic susceptibility is obtained by a semi-analytical approach assuming only the first order magneto-optical coupling \cite{wojewoda_modeling_2024}.

\begin{figure}
\includegraphics{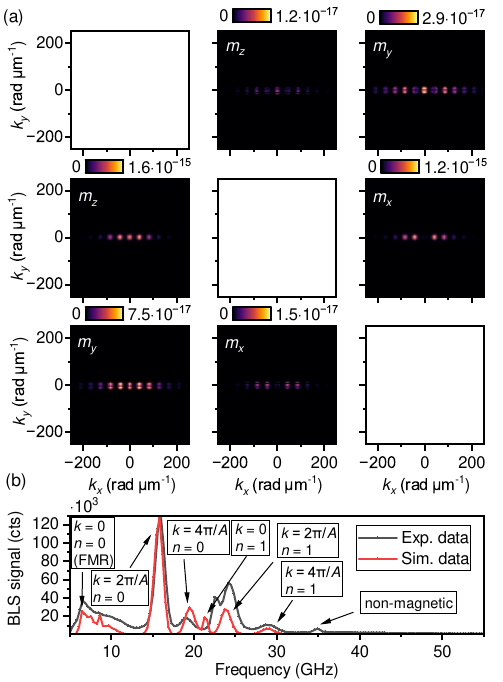}
\caption{\label{fig:Fig3} FDTD simulation data and results for a dielectric array with periodicity $A=150\,$nm. (a) Transfer function tensor components, with each panel labeled by the corresponding dynamic magnetization component that it interacts with (assuming only linear magneto-optical coupling). The diagonal components are not computed, as they don’t contribute to the final signal. (b) Simulated BLS spectra compared with experimental results. Peaks are annotated with their corresponding wavenumbers and perpendicular standing spin wave (PSSW) modes.}
\end{figure}

We apply this model to obtain the transfer function [Fig.~\ref{fig:Fig3}(a)] of silicon nanostripes with the periodicity $A = 150\,$nm, with materials and geometry matching with the fabricated sample. As expected, the transfer function shows maxima at the multipliers of the periodicity, i.e. $2\pi/A=42$\,rad\,{\textmu}m$^{-1}$, $4\pi/A=84$\,rad\,{\textmu}m$^{-1}$. Also, these maxima show strong directionality, as they occur only for the close to zero $y$-wavevector component (along the long axis of the stripes), which is in the agreement with reciprocal distribution of the electric field, see Fig.~\ref{fig:Fig2}(c). This behavior can be exploited to gain directional and magnitude sensitivity in the BLS process.

The transfer matrix components $T_{xy}$ and $T_{zy}$, which couple to the $m_z$ and $m_x$ magnetization components, respectively, are approx. two orders of magnitude larger than the remaining transfer matrix elements. However, depending on the exact geometry of the incident electric field, the direction of the static magnetization, the coupling mechanism and the ellipticity of the spin-wave mode, the absolute contribution to the resulting signal can vary. For example, if the static magnetization is aligned along the $x$-direction and only linear magneto-optical coupling is assumed, the contribution of the $T_{zy}$ component of the transfer matrix function to the resulting signal would be zero.

The width of the illumination spot determines the wavevector resolution (the width of the periodic peaks in the transfer matrix components). The larger the illuminated area, the better wavenumber resolution can be achieved. In our experiment, we have used a 440\,nm wide Gaussian spot, resulting in a wavevector resolution of approx. 10\,rad\,{\textmu}m$^{-1}$. The resolution is independent on the order of the peak of transfer function and on the periodicity of the nanostripe array. The increase in wavevector resolution can only be achieved at the expense of spatial resolution and vice versa.

In Fig.~\ref{fig:Fig3}(b), this transfer function has been used to model the BLS signal and compare it directly with the experiment. We can observe the formation of peaks at the frequencies given by the dispersion relation [see Fig.~\ref{fig:Fig1}(a)] matching with \textit{k}-vector sensitivity imposed by the periodicity of the dielectric stripes ($A=150\,\mathrm{nm}$), see Fig.~\ref{fig:Fig2}(c) and Fig.~\ref{fig:Fig3}(a). The intensity of the \textmu BLS accessible mode on a bare film ($k\lesssim10$\,rad\,{\textmu}m$^{-1}$, $n=0$) is strongly suppressed compared to the peak with matching periodicity to the periodic dielectric nanoresonator array ($k=2\pi/A$, $n=0$). The intensity of the matching higher order peaks ($k=4\pi/A$, $k=6\pi/A$, etc.) decays exponentially, so only the first two orders are visible in this representation. However, thanks to the high contrast of the tandem Fabry-Perot interferometer (150\,dB), even very small contributions can be resolved experimentally. 

At higher frequencies we can also observe peaks caused by the first-order perpendicular standing spin waves (PSSW, $n=1$). There, we can see a small frequency offset between the model and the experimental data, which will be discussed later in the text.

\begin{figure}
\includegraphics{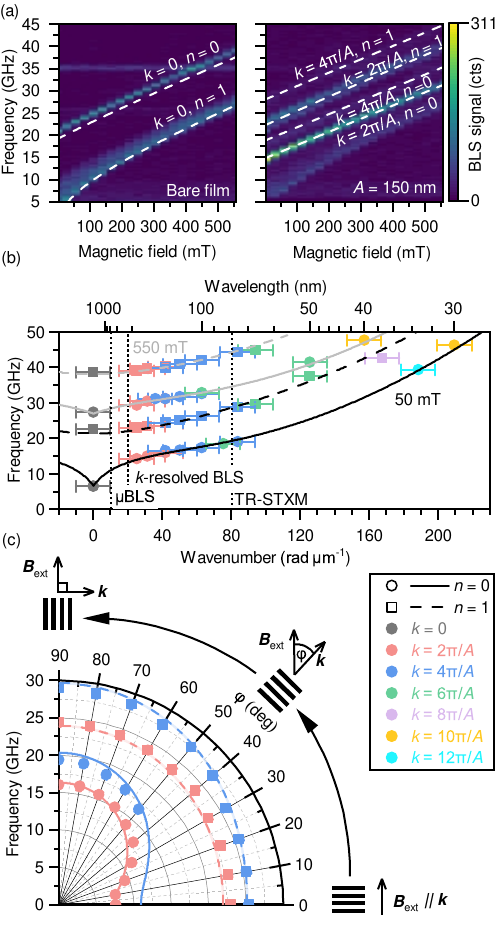}
\caption{\label{fig:Fig4} Experimental results measured on arrays with varying periodicity $A$. (a) Comparison of 2D BLS spectra (frequency vs magnetic field) for a bare magnetic film and a dielectric array. The presence of the array leads to additional accessible magnon wavenumbers and their higher-order thickness modes. The curves represent theoretical calculations using the material constants obtained from (b). (b) Magnon dispersion relation obtained from arrays with periodicities ranging from 75 to 150\,nm, showing discrete modes (data points) fitted with theoretical dispersion curves using the Kalinikos-Slavin model \cite{kalinikos_theory_1986, githubSWT} (c) Angular dependence of the frequency of magnons with specific wavenumber on the angle between the external magnetic field and the magnon wavevector. The polarization of the incident laser light was perpendicular to the magnetic field. The curves represent theoretical predictions calculated using fitted material parameters from (b).}
\end{figure}

To verify the origin of the signal, we measured the BLS spectra on the same dielectric stripes ($A=150\,\mathrm{nm}$) in different external magnetic fields and compared them with a standard \textmu BLS measurement on the bare film, see Fig.~\ref{fig:Fig4}(a). On the bare film [Fig.~\ref{fig:Fig4}(a), left panel], only spin waves around the center of the Brillouin zone ($k=0$) are visible for both PSSW modes. However, as mentioned before, with the periodic dielectric nanostripes, we resolve spin waves with wavenumbers corresponding to the first two multiples of $2\pi/A$. For both experiments, we can clearly observe the expected dependency in field, which agrees well with the analytical formula for spin wave dispersion \cite{kalinikos_theory_1986, githubSWT}.

To extract the dispersion relation of the NiFe film, we performed a similar experiment as in Fig.~\ref{fig:Fig3}(b). We used multiple nanostripe arrays with periodicities ranging from 150 to 300\,nm and measured BLS signal in low (50\,mT) and high (550\,mT) magnetic fields. By analyzing the measured peak positions, we were able to reconstruct the dispersion relation [Fig.~\ref{fig:Fig4}(b)]. With just eight separate measurements, we were able to obtain a wide wavenumber range spanning from dipolar to exchange dominated spin waves. The experimentally obtained results were fitted with theoretical dispersion relations \cite{kalinikos_theory_1986, githubSWT}. The wide range of measured wavenumbers allowed us to extract from a single fit the saturation magnetization $M_{\mathrm{s}}=(700\pm20)\,\mathrm{kA/m}$, the gyromagnetic ratio $\gamma/2\pi=(30.4\pm0.4)\,\mathrm{GHz/T}$, and also usually strongly correlated material parameters -- the exchange constant $A_{\mathrm{ex}}=(8.1\pm0.3)\,\mathrm{pJ/m}$ and the film thickness $t=(26.0\pm0.3)\,\mathrm{nm}$. However, the model has a frequency offset (approx. $1.5$\,GHz) from the experimental data for modes $n=1$, $k=0$. This may be due to more complicated boundary conditions than totally unpinned used for the fit.

To demonstrate the directional sensitivity of the BLS process on periodic dielectric  nanostripes, we rotated the sample along the $z$-axis while keeping the direction of the external magnetic field and the incident electric field constant, and acquired the BLS spectra. For each angle, the frequency positions of magnon modes with specific wavenumbers were extracted [Fig.~\ref{fig:Fig4}(c)]. Regarding the fundamental spin wave mode ($n=0$, solid line) we can observe a strong anisotropy in the dispersion relation caused by dipolar interaction. For the first PSSW ($n=1$, dashed line), where the dipolar interaction is much weaker compared to the exchange interaction, the spin wave frequencies are almost isotropic. The obtained data agree well with the analytical models \cite{kalinikos_theory_1986, githubSWT}.


In conclusion, we have demonstrated a novel approach for measuring magnons in previously inaccessible wavevector range using Mie-enhanced micro-focused Brillouin light scattering. By integrating periodic dielectric nanoresonators into a standard \textmu BLS experiment, we achieved full wavevector resolution, enabling the detection of magnons with wavelengths down to 30\,nm (wavevectors up to 200\,rad\,{\textmu}m$^{-1}$). This technique bridges a long-standing experimental gap, providing new opportunities for investigating fundamental spin-wave physics across the entire Brillouin zone or for imaging nanoscale spin-wave devices. Our results establish a robust method for characterizing nanoscale magnons in space, wavenumber, and frequency with unprecedented resolution, validated through both experimental measurements and theoretical modeling.

Beyond magnonics, the methodology developed here is also applicable to other collective excitations, such as phonons, potentially opening new avenues in fields ranging from ultrafast magnetism to mechanobiology \cite{choi2014spin, prevedel_brillouin_2019}. The ability to resolve excitations with nanoscale precision and directional sensitivity could further drive advancements in the scope of wavevector-resolved spectroscopy in condensed matter physics.

\begin{acknowledgements}
We thank Burkard Hillebrands for insightful discussions. This research was supported by the project No. CZ.02.01.01\slash00\slash22\textunderscore008\slash0004594 (TERAFIT) and by the Grant Agency of the Czech Republic, project no. 23-04120L. CzechNanoLab project LM2023051 is acknowledged for the financial support of the measurements and sample fabrication at CEITEC Nano Research Infrastructure.
\end{acknowledgements}

\section*{Data statement}
All data and code used to generate the presented figures are available in the Zenodo repository (DOI: 10.5281/zenodo.14801809).

\bibliography{apssamp}

\section*{End matter}
\subsection*{Finite-difference time-domain simulations}
Simulations were conducted using Ansys Lumerical FDTD Solutions software. The simulation region measured $8000\times8000\times1150\,\mathrm{nm}^3$, with the shortest dimension aligned along the optical axis. The model consisted of a semi-infinite silicon substrate coated with a 27.6\,nm-thick permalloy film. On top, amorphous silicon nanostripes with a periodicity of 150\,nm were arranged, covering the entire simulation area. Amorphous silicon was selected due to its closer optical resemblance to sputtered silicon.

The global mesh was set to conformal variant 0 (mesh order 4), with a refined 2\,nm mesh in the central simulation region ($1024\times1024\times96\,\mathrm{nm}^3$) in the vicinity of the nanostripes, approximately corresponding to the laser spot position. Perfectly matched layers were used as boundary conditions, and appropriate symmetry conditions were applied. A Gaussian source, modeled using the thin-lens approximation with a numerical aperture of 0.75, was focused on the top of the center stripe. The source polarization was set either perpendicular or parallel to the nanostripes.

The dielectric functions for the silicon substrate, amorphous silicon, and permalloy were obtained from \cite{palik1998, PhysRevB.5.3017, Tikuisis2017}. Electric field data were collected using a field monitor positioned in the middle of the permalloy film and analyzed with MATLAB 2024a.

\subsection*{BLS spectra modeling using the reciprocity theorem of electromagnetism}
\begin{figure}[ht!]
\includegraphics{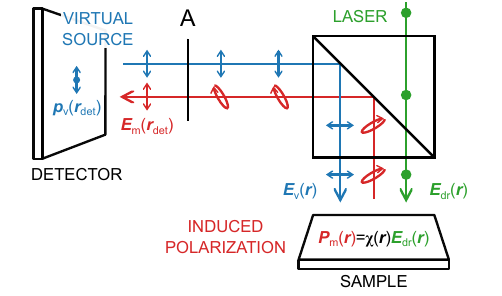}
\caption{\label{fig:EM-FIG2} Illustration of the reciprocity theorem concept. It highlights the key physical quantities and their polarization states – parallel, perpendicular, and arbitrary (elliptic). The letter A denotes an analyzer.}
\end{figure}

The task of calculating the intensity of the BLS signal collected by a detector can be elegantly completed utilizing the reciprocity theorem \cite{neuman_mapping_2015}, where we exploit a close link between the fields generated by an electric dipole ($\boldsymbol{p}$) (emulating the polarization current induced within the magnetic layer) and a virtual source located at the position of the detector, see Fig.~\ref{fig:EM-FIG2}. Mathematically, this connection can be written down as

\begin{equation}
\label{eq:ReciTheorem1}
\boldsymbol{p}_{\mathrm{v} } (\boldsymbol{r}_{\mathrm{det} }) \cdot \boldsymbol{E}_{\mathrm{m} } (\boldsymbol{r}_{\mathrm{det} }) = \boldsymbol{p}_{\mathrm{m} } (\boldsymbol{r}) \cdot \boldsymbol{E}_{\mathrm{v} } (\boldsymbol{r}),
\end{equation}

\noindent
where the subscripts differentiate whether the particular field or dipole moment is associated with the magnetic source (m) or the virtual source (v). Note that the two dot products are evaluated at different positions in space corresponding to the respective locations of the sources (detector and magnetic sample). The above relation can be translated into a simple statement: if a photon emitted from the detector is able to reach a particular spot on the sample, then it should work also the other way around. Although this is not true in general (e.g. in gyrotropic materials \cite{Potton_2004}), the materials in our sample are linear and isotropic (at least with respect to the out-coupling process that follows the inelastic scattering event) and the reciprocity theorem is valid. While the amplitude of the virtual dipole source can be set arbitrarily, its orientation should match the experimental conditions under which the BLS signal is collected. In presented experiments and calculations, we employ a cross-polarized scheme, which is sensitive to the spin wave BLS signal in linear case \cite{wojewoda_modeling_2024}. Most importantly, the virtual electric field should capture all the effects arising from the propagation between the detector and the sample, including any interaction with the silicon stripes or surrounding media. We assume that the collection and illumination spots are identical with a width of 440\,nm and a Gaussian shape \cite{wojewoda_observing_2023}. This is justified, because both branches share the same objective lens and are perfectly aligned. Taking all this information as new settings for our FDTD model, we calculated the distribution of the virtual electric field within the magnetic layer.

To account for the continuous nature of the induced polarization $\boldsymbol{P}_{\mathrm{m} } (\boldsymbol{r}) = \boldsymbol{\chi} (\boldsymbol{r}, \omega_{\mathrm{m} } ) \boldsymbol{E}_{\mathrm{dr} } (\boldsymbol{r})$ (implicitly assuming linear magneto-optical coupling), the right-hand side of Eq.~\ref{eq:ReciTheorem1} should be replaced by a volume integral over the magnetic layer 

\begin{align}
\label{eq:ReciTheorem2}
\boldsymbol{p}_{\mathrm{v} } (\boldsymbol{r}_{\mathrm{det} }) & \cdot \boldsymbol{E}_{\mathrm{m} } (\boldsymbol{r}_{\mathrm{det} })  = \int \mathrm{d}r^{3} \, \boldsymbol{P}_{\mathrm{m} } (\boldsymbol{r}) \cdot \boldsymbol{E}_{\mathrm{v} } (\boldsymbol{r}) = \notag \\
& = \sum\limits_{i,j} \int \mathrm{d}r^{3} \, {\chi}_{\mathrm{m} }^{ij} (\boldsymbol{r}) \, {E}_{\mathrm{v} }^{i} (\boldsymbol{r}) {E}_{\mathrm{dr} }^{j} (\boldsymbol{r})
\end{align}

\noindent
Restricting our analysis to laterally propagating magnons, the magnetic susceptibility tensor can be cast as

\begin{equation}
\label{eq:ReciTheorem3}
\boldsymbol{\chi}_{\mathrm{m} } (\boldsymbol{r}, \omega_{\mathrm{m} }) = \boldsymbol{\chi}_{\mathrm{m} } (z,\omega_{\mathrm{m} }) \, e^{i \boldsymbol{k}_{\mathrm{m} } \cdot \boldsymbol{r}_{\parallel} },
\end{equation}

\noindent
where $\boldsymbol{k}_{\mathrm{m} }$ denotes the magnon wavevector and the symbol ${\parallel}$ indicates the lateral (in-plane of the magnetic sample) spatial coordinates. The variation of the susceptibility tensor with respect to the vertical coordinate $z$ will generally depend on the thickness of the magnetic layer and the PSSW order. Since we are studying metallic layers (NiFe), where the penetration depth of light ($\approx 20$\,nm) is comparable or shorter than the thickness of the sample, we neglect this dependence in our analysis. Inserting Eq.~\ref{eq:ReciTheorem3} into Eq.~\ref{eq:ReciTheorem2} and recalling that the detected BLS signal is proportional to the intensity of the collected light, we find the contribution from a single magnon with a wavevector $\boldsymbol{k}_{\mathrm{m} }$ to be

\begin{align}
\label{eq:ReciTheorem4}
\sigma( & \boldsymbol{k}_{\mathrm{m} }, \omega_{\mathrm{m}} )  \sim \vert \boldsymbol{p}_{\mathrm{v} } (\boldsymbol{r}_{\mathrm{det} }) \cdot \boldsymbol{E}_{\mathrm{m} } (\boldsymbol{r}_{\mathrm{det} }) \vert^{2} = \notag \\ & = \Big\vert \sum\limits_{i,j} \int \mathrm{d}z \, {\chi}_{\mathrm{m} }^{ij} (z,\omega_{\mathrm{m} }) \int \mathrm{d}r_{\parallel}^{2} \, {E}_{\mathrm{v} }^{i} (\boldsymbol{r}) {E}_{\mathrm{dr} }^{j} (\boldsymbol{r}) \, e^{i \boldsymbol{k}_{\mathrm{m} } \cdot \boldsymbol{r}_{\parallel} } \Big\vert^{2}.
\end{align}

\noindent
The above equation indicates that the strength of the BLS signal is largely determined by the overlap integral between the driving and virtual electric field distributions modulated at the frequency of the magnon. It can be perceived as a transfer function of the system that determines the range of detectable magnons and its inspection can be of great value for both the design process and the subsequent analysis of experimental measurements.

\subsection*{Thickness determination of the permalloy film via XRR}

\begin{figure}[bt]
\includegraphics{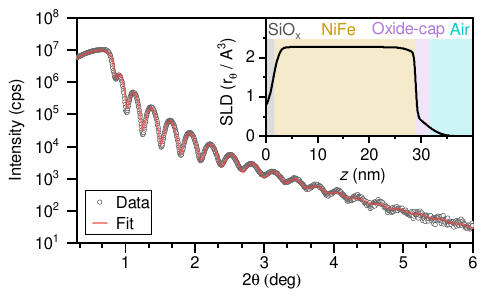}
\caption{\label{fig:EM-FIG1} Measured XRR data (open symbols) and corresponding fit (solid line) for the permalloy layer used in the study. A Si(substrate)/$\mathrm{SiO}_x$/NiFe/oxide-cap multilayer model was employed for the thickness determination of the layers. The inset shows the depth-dependence of the scattering length density (SLD) resulting from the model refinement,  where $r_{\theta}$ is the classical electron radius and $z = 0$ corresponds to the Si substrate height.}
\end{figure}

The thickness of the permalloy film was determined via X-ray reflectivity (XRR) measurements using a Rigaku Smartlab (9\,kW) diffractometer with Cu-K$\alpha$ radiation ($\lambda = 1.54\,\text{\AA}$) and an incident parallel beam setting. Soller slits of $5^{\circ}$ in both the incident and diffracted optics were employed. The measured intensity vs. $2\theta$ profile was fitted using the GenX 3 software \cite{glavic_genx_2022} (see Fig.~\ref{fig:EM-FIG1}). The multilayer refinement provides for the permalloy film fitted thickness and density values of $27.6\pm0.1\,\mathrm{nm}$ and $8.7\pm 0.1\,\mathrm{g/cm}^3$, respectively. A thin capping layer with a reduced density ($\sim30\,\%$ of the permalloy density value) in the model considers the naturally formed oxide on top of the magnetic layer, for which a thickness of $2.5\pm0.5\,\mathrm{nm}$ is obtained.

\end{document}